\documentclass[12pt]{article}
\usepackage{wrapfig}
\usepackage{epsfig}
\usepackage{hyperref}
\usepackage{amssymb}
\usepackage{amsmath,mathtools}
\usepackage{amsfonts}
\usepackage{latexsym}
\usepackage{wasysym}
\usepackage{multirow}
\usepackage{fixmath}
\usepackage{color}
\usepackage{stackrel}
\usepackage{txfonts}                                                            \usepackage{centernot}
\usepackage{bbm}
\usepackage[low-sup]{subdepth}
\usepackage{hyperref}

\newcommand{\bea}{\begin{eqnarray}}
\newcommand{\eea}{\end{eqnarray}}
\newcommand{\bite}{\begin{itemize}}
\newcommand{\eite}{\end{itemize}}

\date{}

\begin{document}
\title{
\vspace{-2.0cm} 
\flushleft
{\normalsize ADP-17-09/T1015} \\
\vspace{-0.35cm}
{\normalsize DESY 17-027} \\
\vspace{-0.35cm}
{\normalsize Edinburgh 2017/04} \\
\vspace{-0.35cm}
{\normalsize Liverpool LTH 1122} \\
\vspace{-0.35cm}
{\normalsize March 2017} \\
\vspace{0.5cm}
\centering{\bf Nucleon structure functions from lattice operator product expansion}}

\author{\large G. Schierholz}
\author{A.J.~Chambers$^a$, R.~Horsley$^b$, Y.~Nakamura$^c$, H.~Perlt$^d$, P.E.L.~Rakow$^e$,\\ G.~Schierholz$^f$, A.~Schiller$^d$, K.~Somfleth$^a$, R.D.~Young$^a$ and J.M.~Zanotti$^a$\\[1em] 
$^a$ CSSM, Department of Physics, University of Adelaide,\\ Adelaide SA 5005, Australia\\[0.15em]
$^b$ School of Physics and Astronomy, University of Edinburgh,\\ Edinburgh
EH9 3FD, United Kingdom\\[0.15em] 
$^c$ RIKEN Advanced Institute for Computational Science,\\ Kobe, Hyogo 650-0047, Japan\\[0.15em] 
$^d$ Institut f\"ur Theoretische Physik, Universit\"at Leipzig,\\ 04103
Leipzig, Germany\\[0.15em]  
$^e$ Theoretical Physics Division, Department of Mathematical Sciences,\\
University of Liverpool, Liverpool L69 3BX, United Kingdom\\[0.15em] 
$^f$ Deutsches Elektronen-Synchrotron DESY,\\ 22603 Hamburg, Germany}

\maketitle
\vspace*{-0.5cm}

\begin{center}
{\large -- QCDSF Collaboration --}
\end{center}

\begin{abstract}
Deep-inelastic scattering, in the laboratory and on the lattice, is most instructive for understanding how the nucleon is built from quarks and gluons. The long-term goal is to compute the associated structure functions from first principles. So far this has been limited to model calculations. In this Letter we propose a new method to compute the structure functions directly from the virtual, all-encompassing Compton amplitude, utilizing the operator product expansion. This overcomes issues of renormalization and operator mixing, which so far have hindered lattice calculations of power corrections and higher moments.
\end{abstract}

\newpage

The connection between the deep-inelastic structure functions and the quark structure of the nucleon is commonly rendered by the parton model. Although providing an intuitive language, in which to interpret the deep-inelastic scattering data, the parton model is incomplete. The theoretical basis is the operator product expansion (OPE). The operators in the expansion are classified according to twist. The parton model accounts for twist-two contributions only, and cannot accommodate power corrections arising from operators of higher twist. Power corrections are inseparably connected with the leading-twist contributions, as a result of operator mixing~\cite{Martinelli:1996pk}. 
Consider, for example, a generic moment of any deep-inelastic structure function of the nucleon regularized on the lattice by a hard cut-off $1/a$, 
\begin{equation}
\label{op2}
\mu(q^2) = c_2(q^2a^2)\, v_2(a) + \frac{c_4(q^2a^2)}{q^2}\,v_4(a) + \cdots \,,
\end{equation}
where $v_2=\langle N|\mathcal{O}_2(a)|N\rangle$ and $v_4=\langle N|\mathcal{O}_4(a)|N\rangle$ are reduced nucleon matrix elements of local operators of twist two and four, respectively, and $c_2$ and $c_4$ are the corresponding reduced Wilson coefficients. The operator $\mathcal{O}_4$ mixes with $\mathcal{O}_2$ with mixing coefficients which diverge as $1/a^2$. The power divergences of $v_4(a)$ must be cancelled with those of $c_2(q^2a^2)$, which demands a nonperturbative calculation of the Wilson coefficient as well.
This can be accomplished by an entirely nonperturbative calculation of the structure functions only.

So far lattice calculations of nucleon structure functions have been limited to calculations on the parton level~\cite{dislat}. In this Letter we propose a method that goes beyond that and computes the deep-inelastic structure functions, including power corrections, directly from the product of electromagnetic currents. This approach has been called `OPE without OPE' elsewhere~\cite{Martinelli:1998hz}. For previous and related work on the subject see~\cite{op,Caracciolo,Detmold:2005gg}.


The starting point is the forward Compton amplitude of the nucleon~\cite{dis},
\begin{equation}
T_{\mu\nu}(p,q) = \rho_{\lambda \lambda^\prime}\! \int\! {\rm d}^4\!x\, {\rm e}^{iq\cdot x}  \langle p,\lambda^\prime |T J_\mu(x) J_\nu(0)|p,\lambda\rangle \,, 
\label{prod}
\end{equation}
the time ordered product of electromagnetic currents sandwiched between nucleon states of momentum $p$ and polarization $\lambda$, where $q$ is the momentum of the virtual photon and $\rho$ is the polarization density matrix. For simplicity, we will restrict ourselves to unpolarized structure functions only with $2 \rho=\mathbf{1}$. We are then left with
\begin{equation}
T_{\mu\nu}(p,q) = \left(\delta_{\mu\nu}-\frac{q_\mu q_\nu}{q^2}\right) \mathcal{F}_1(\omega,q^2) + \left(p_\mu-\frac{p\cdot q}{q^2}q_\mu\right) \left(p_\nu-\frac{p\cdot q}{q^2}q_\nu\right) \frac{1}{p\cdot q} \mathcal{F}_2(\omega,q^2) \,,
\label{dec}
\end{equation}
where $\omega = 2p\cdot q/q^2$. Euclidean metric is understood. Crossing symmetry, $T_{\mu\nu}(p,q)=T_{\nu\mu}(p,-q)$, implies that $\mathcal{F}_1$ is an even function of $\omega$ and $\mathcal{F}_2$ an odd function,
$\displaystyle \mathcal{F}_1(-\omega,q^2) = \mathcal{F}_1(\omega,q^2)$, $\displaystyle \mathcal{F}_2(-\omega,q^2) = -\mathcal{F}_2(\omega,q^2)$.
In the physical region $1 \leq |\omega| \leq \infty$
\begin{equation}
{\rm Im}\, \mathcal{F}_1(\omega,q^2) = 2\pi F_1(\omega,q^2)\,, \quad {\rm Im}\, \mathcal{F}_2(\omega,q^2) = 2\pi F_2(\omega,q^2)\,,
\label{cross}
\end{equation}
where $F_1$ and $F_2$ are the deep-inelastic structure functions of the nucleon.
Using the OPE, one can express $\mathcal{F}_1$ and $\mathcal{F}_2$ in terms of moments of $F_1$ and $F_2$, which are amenable to calculation on the Euclidean lattice. Alternatively, $\mathcal{F}_1$ and $\mathcal{F}_2$ can be written as dispersion integrals over $\omega$, which leads to the same expressions.

Let us first consider the OPE of $\mathcal{F}_1$ and $\mathcal{F}_2$. After some simple algebra we obtain~\cite{dis}
\begin{equation}
\begin{split}
T_{\mu\nu}(p,q) = \sum_{n=2,4,\cdots}^\infty &\left\{ \left(\delta_{\mu\nu}-\frac{q_\mu q_\nu}{q^2}\right)\, 4\omega^n \int_0^1 dx\, x^{n-1} F_1(x,q^2)\right. \\[0.5em]
&\!\!\left.+\left(p_\mu-\frac{p\cdot q}{q^2}q_\mu\right)\left(p_\nu-\frac{p\cdot q}{q^2}q_\nu\right)\, \frac{8}{2p\cdot q}\, \omega^{n-1} \int_0^1 dx\, x^{n-2} F_2(x,q^2)\right\} \,.
\end{split}
\label{opes1}
\end{equation}
The series $\sum_{k \in \mathbf{N}}\, (\omega x)^{2k}$ in (\ref{opes1}) is geometric and sums up to $\displaystyle [1-(\omega x)^2]^{-1}$, which leads to the alternate expression
\begin{equation}
\begin{split}
T_{\mu\nu}(p,q) &= \left(\delta_{\mu\nu}-\frac{q_\mu q_\nu}{q^2}\right)\, 4\omega \int_0^1 dx\, \frac{\omega x}{1-(\omega x)^2}\, F_1(x,q^2) \\[0.5em]
&\,+ \left(p_\mu-\frac{p\cdot q}{q^2}q_\mu\right)\left(p_\nu-\frac{p\cdot q}{q^2}q_\nu\right)\, \frac{8\omega}{2p\cdot q} \int_0^1 dx\, \frac{1}{1-(\omega x)^2}\, F_2(x,q^2) \,.
\end{split}
\label{opes2}
\end{equation}
In the limit where $F_1(x,q^2)$ and $F_2(x,q^2)$ become independent of $q^2$ we have the Callan-Gross relation $\displaystyle F_2(x) = 2 x F_1(x)$.

Alternatively, we can express $\mathcal{F}_1$ and $\mathcal{F}_2$ directly in terms of the structure functions $F_1$ and $F_2$, circumventing the OPE. The amplitudes $\mathcal{F}_1$ and $\mathcal{F}_2$ have cuts at $-\infty \leq \omega \leq - 1$ and $1 \leq \omega \leq \infty$ with discontinuities (\ref{cross}). This leads to once subtracted dispersion relations
\begin{equation}
\begin{split}
\mathcal{F}_1(\omega,q^2) &= 2 \omega \int_1^\infty d\bar{\omega} \left[\frac{F_1(\bar{\omega},q^2)}{\bar{\omega}\,(\bar{\omega}-\omega)}-\frac{F_1(\bar{\omega},q^2)}{\bar{\omega}\,(\bar{\omega}+\omega)}\right] + \mathcal{F}_1(0,q^2)\,,\\[0.5em]
\mathcal{F}_2(\omega,q^2) &= 2 \omega \int_1^\infty d\bar{\omega} \left[\frac{F_2(\bar{\omega},q^2)}{\bar{\omega}\,(\bar{\omega}-\omega)}+\frac{F_2(\bar{\omega},q^2)}{\bar{\omega}\,(\bar{\omega}+\omega)}\right] \,.
\end{split}
\label{pola}
\end{equation}
While $\mathcal{F}_2(0,q^2)=0$, the subtraction constant $\mathcal{F}_1(0,q^2)$ contains information on the magnetic polarizability of the nucleon and the proton--neutron electromagnetic mass shift~\cite{Agadjanov:2016cjc}. In the following equations we shall discard it, as it has no counterpart in $F_1$, nor is it accounted for by the OPE. It can be computed like any other value of $\mathcal{F}_1$ though and, if necessary, has to be subtracted from $\mathcal{F}_1(\omega,q^2)$. (So, for example, from the data underlying Fig.~\ref{fig6}.) Substituting $\bar{\omega}$ by $1/x$, we finally obtain
\begin{equation}
\mathcal{F}_1(\omega,q^2) = 4 \omega^2 \int_0^1 dx\, x \, \frac{F_1(x,q^2)}{1-(\omega x)^2} \,, \quad
\mathcal{F}_2(\omega,q^2) = 4 \omega \int_0^1 dx\, \frac{F_2(x,q^2)}{1-(\omega x)^2}\,,
\label{disp}
\end{equation}
where we have identified $F_1(\bar{\omega},q^2)$ and $F_2(\bar{\omega},q^2)$ with $F_1(x,q^2)$ and $F_2(x,q^2)$, respectively. If we insert (\ref{disp}) into (\ref{dec}), we obtain (\ref{opes2}), in agreement with the OPE resummed. It should be noted that the structure functions $F_1(x,q^2)$ and $F_2(x,q^2)$ include higher twist contributions, as we have not made any assumptions on $\mathcal{F}_1$ and $\mathcal{F}_2$ other than on the analytic structure.

\begin{figure}[t]
\begin{center}
\epsfig{file=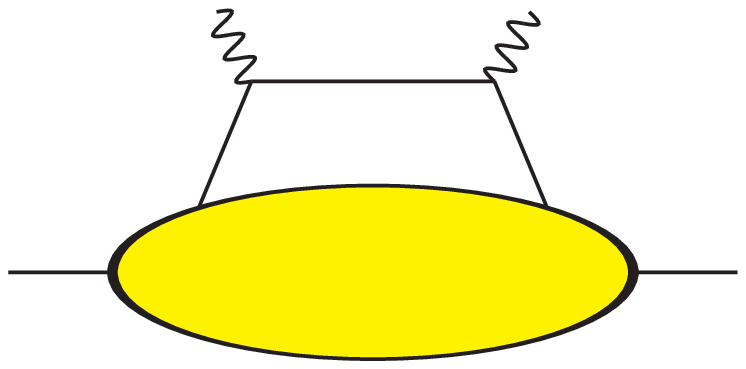,width=7.5cm,clip=} \hspace*{-1cm}
\epsfig{file=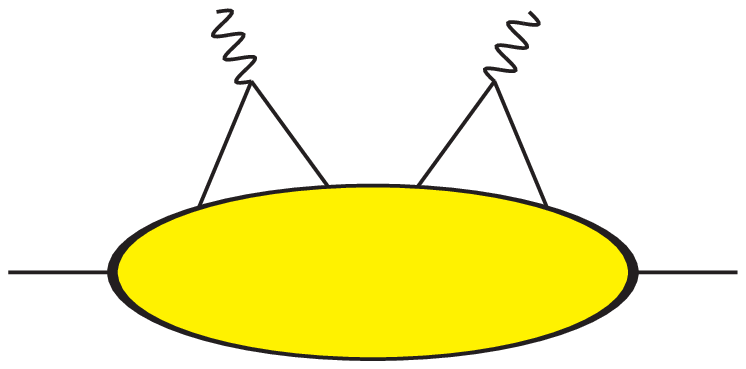,width=7.5cm,clip=}
\end{center}
\vspace*{-0.75cm}
\caption{The so called `handbag' diagram (left panel) and `cats-ears' diagram (right panel).}
\label{fig1}
\end{figure}

To simplify the numerical calculation, we may choose $\mu = \nu = 3$ and $p_3 = q_3 = q_4 = 0$. We then have
\begin{equation}
T_{33}(p,q) = \sum_{n=2,4,\cdots}^\infty 4\omega^n \int_0^1 dx\, x^{n-1} F_1(x,q^2) 
\label{opess1}
\end{equation}
and, alternatively, 
\begin{equation}
T_{33}(p,q) =  4\omega \int_0^1 dx\, \frac{\omega x}{1-(\omega x)^2} F_1(x,q^2) \,. 
\label{opess2}
\end{equation}
For $|\omega|>1$ the principal value has to be taken. The matrix element $T_{33}(p,q)$ can be computed most efficiently, including singlet matrix elements, by a simple extension of existing implementations of the Feynman-Hellmann technique to lattice QCD~\cite{Horsley:2012pz}. For simplicity, we consider the local vector current only. The appropriate renormalization factor $Z_V$ can be computed unambiguously~\cite{Bakeyev:2003ff}. No further renormalization is needed. To compute the Compton amplitude from the Feynman-Hellmann relation, we introduce the perturbation to the Lagrangian
\begin{equation}
\mathcal{L}(x) \rightarrow \mathcal{L}(x) + \lambda \mathcal{J}_3(x)\,, \quad \mathcal{J}_3(x)=Z_V\cos(\vec{q}\cdot\vec{x})\; e_f \,\bar{\psi}_f(x)\gamma_3 \psi_f(x) \,,
\label{add}
\end{equation}
where $\psi_f$ is the quark field of flavor $f=u, d, s, \cdots$ to which the photon is attached, and $e_f$ is its electric charge. Note that $\lambda$ has dimension mass. 
%
Taking the second derivative of the nucleon two-point function $\displaystyle \langle N(\vec{p},t) \bar{N}(\vec{p},0)\rangle_\lambda \simeq C_\lambda\, {\rm e}^{-E_\lambda(p,q)\,t}$ with respect to $\lambda$ on both sides, we obtain 
\begin{equation}
-2 E_\lambda(p,q)\, \frac{\partial^2}{\partial\lambda^2}  E_\lambda(p,q)\,\big|_{\lambda=0} = T_{33}(p,q)\,.
\label{fh}
\end{equation}
The derivation of (\ref{fh}) would go beyond the scope of this Letter and will be presented in a separate publication. Provided we compute at sufficiently large $q^2$, standard factorization theorems state that the Compton amplitude will be dominated by the `handbag' diagram shown in the left panel of Fig.~\ref{fig1}. Nevertheless, the amplitude does encompass all contributions, including the power-suppressed `cats-ears' diagram shown in the right panel. Thereby, varying $q^2$ will allow us to test the twist expansion and, in particular, isolate twist-four contributions.  
A conventional calculation of the four-point function $\langle p,\lambda |J_\mu(x) J_\nu(0)|p,\lambda\rangle$, in contrast, would involve all-to-all quark propagators twice consecutively, which is not practicable.

%


Knowing $T_{33}(p,q)$ for sufficiently many values of $\omega$, we can solve (\ref{opess1}) for the moments $\mu_n = \int_0^1 dx x^n F_1(x,q^2)$ and (\ref{opess2}) directly for $F_1(x,q^2)$. We want to keep the photon momentum $\vec{q}$ fixed, but vary the nucleon momentum $\vec{p}$. This amounts to a relatively cheap calculation, as all nucleon energies $E_\lambda(p,q)$ can be computed from a single set of background field configurations (per value of $\lambda$). For nonsinglet quantities, in which the currents couple to valence quarks only, no additional gauge field configurations will have to be generated at all. 



Regarding (\ref{opess1}), the task is to compute (say) the lowest $M$ moments $\mu$ from a finite number of sampled points 
\begin{equation}
t_i=T_{33}(\omega_i) \,,\; i=1, \cdots, N 
\label{in}
\end{equation}
with $M \leq N$. For ease of writing we have dropped the dependence on $q^2$. Under these conditions the OPE of $T_{33}$ can be written as a set of equations
\begin{equation}
\left(\begin{tabular}{c}
$t_1$\\
$t_2$\\
$\vdots$\\ 
$t_N$
\end{tabular}\right)\, =\,
\left(\begin{tabular}{cccc}
$4\omega_1^2$ & $4\omega_1^4$ & $\cdots$ & $4\omega_1^{2M}$\\
$4\omega_2^2$ & $4\omega_2^4$ & $\cdots$ & $4\omega_2^{2M}$\\
$\vdots$ & $\vdots$ & $\vdots$ & $\vdots$\\ 
\;$4\omega_N^2$ & \;$4\omega_N^4$ & $\cdots$ & $4\omega_N^{2M}$
\end{tabular}\right) \;
\left(\hspace*{-0.5cm}\begin{tabular}{c}
$\mu_1$\\
$\mu_3$\\
$\vdots$\\ 
\hspace*{0.5cm}$\mu_{2M-1}$
\end{tabular}\right) \,.
\label{va}
\end{equation}
The matrix is known as Vandermonde matrix. Efficient algorithms for solving (\ref{va}) can be found in the literature~\cite{nr}. Alternatively, we can cast the $T_{33}(\omega_i)$'s into a simple functional form by the interpolating polynomial
\begin{equation}
T_{33}(\omega) = 4\, \big(\omega^2 \mu_1 + \omega^4 \mu_3 + \cdots + \omega^{2M} \mu_{2M-1}\big) \,. 
\label{poly}
\end{equation} 
The moments $\mu$ can then be determined from a fit of (\ref{poly}) to the sampled points (\ref{in}).

To solve (\ref{opess2}), we approximate the integral by a sum over a discrete set of $M$ points, $0 < x_1 < x_2 < \cdots < x_M < 1$, and write
\begin{equation}
f_i = F_1(x_i) \,, \quad K_{ij} = \frac{4\,\omega_i^2x_j}{1-(\omega_i x_j)^2} \,.
\end{equation}
We assume the points to be equidistant with step size $\epsilon$. A generalization to adaptive step sizes is straightforward. The integral equation (\ref{opess2}) then reduces to the set of equations 
\begin{equation}
t_i = \epsilon \sum_{j=1}^M K_{ij}\, f_j \,,\; i=1, \cdots, N \,.
\label{ieq}
\end{equation}
In general, $N < M$. The $N \times M$ matrix $K$ can be written as the product of a $N \times N$ orthogonal matrix $U$, a $N \times N$ diagonal matrix $W$ with positive or zero eigenvalues $w_1 < w_2 < \cdots < w_N$, and the transpose of a row-orthogonal $M \times N$ matrix $V$,
\begin{equation}
K = U \,\left[{\rm diag}(w_1, \cdots, w_N)\right]\,V^T \,.
\end{equation}
The matrix $W$ is singular. Singular value decomposition (SVD) is the method of choice for solving (\ref{opess2}) under such conditions. The solution is
\begin{equation}
f_j = \sum_{i=1}^N K^{-1}_{ji}\epsilon^{-1}\, t_i \,, 
\label{svd}
\end{equation}
where $K^{-1}$ is the pseudoinverse
\begin{equation}
K^{-1} = V \,\left[{\rm diag}(1/w_1, \cdots, 1/w_L, 0, \cdots, 0)\right]\, U^T 
\end{equation}
with $1/w_l$ being replaced by zero if $w_l=0$, which is assumed for $L < l \leq N$. One has to excercise some discretion at deciding at what threshold to set $1/w_l$ to zero. Several routines, such as {\tt PseudoInverse} of {\it Mathematica}~\cite{ma}, solve this problem automatically.

\begin{figure}[!t]
\begin{center}
\epsfig{file=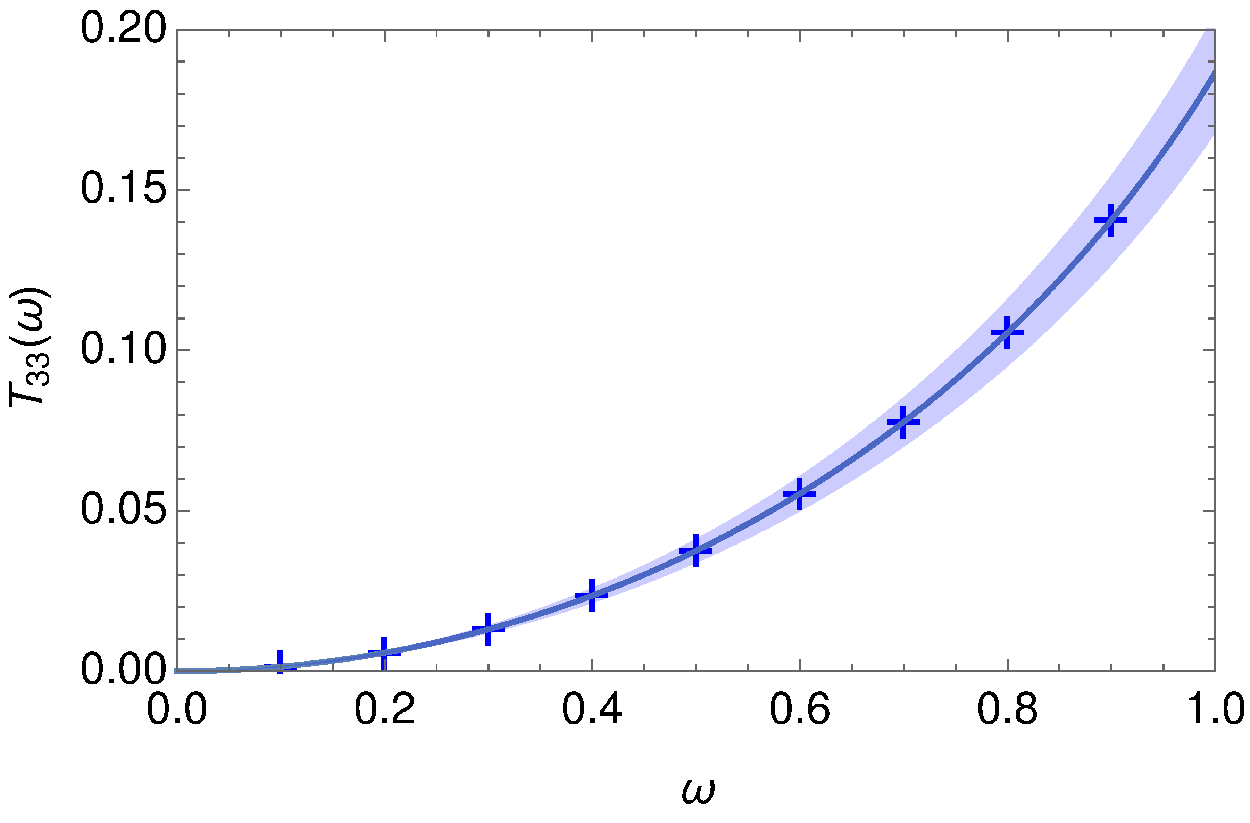,width=10cm,clip=}
\end{center}
\vspace*{-0.75cm}
\caption{The Compton amplitude. The solid curve is the result of (\ref{opess2}) with $F_1^{u-d}$ given by (\ref{mstw}), the pluses are our  `data' points. The shaded area indicates a $10\%$ overall error.}
\label{fig2}
\end{figure}

\begin{figure}[!b]
\begin{center}
\epsfig{file=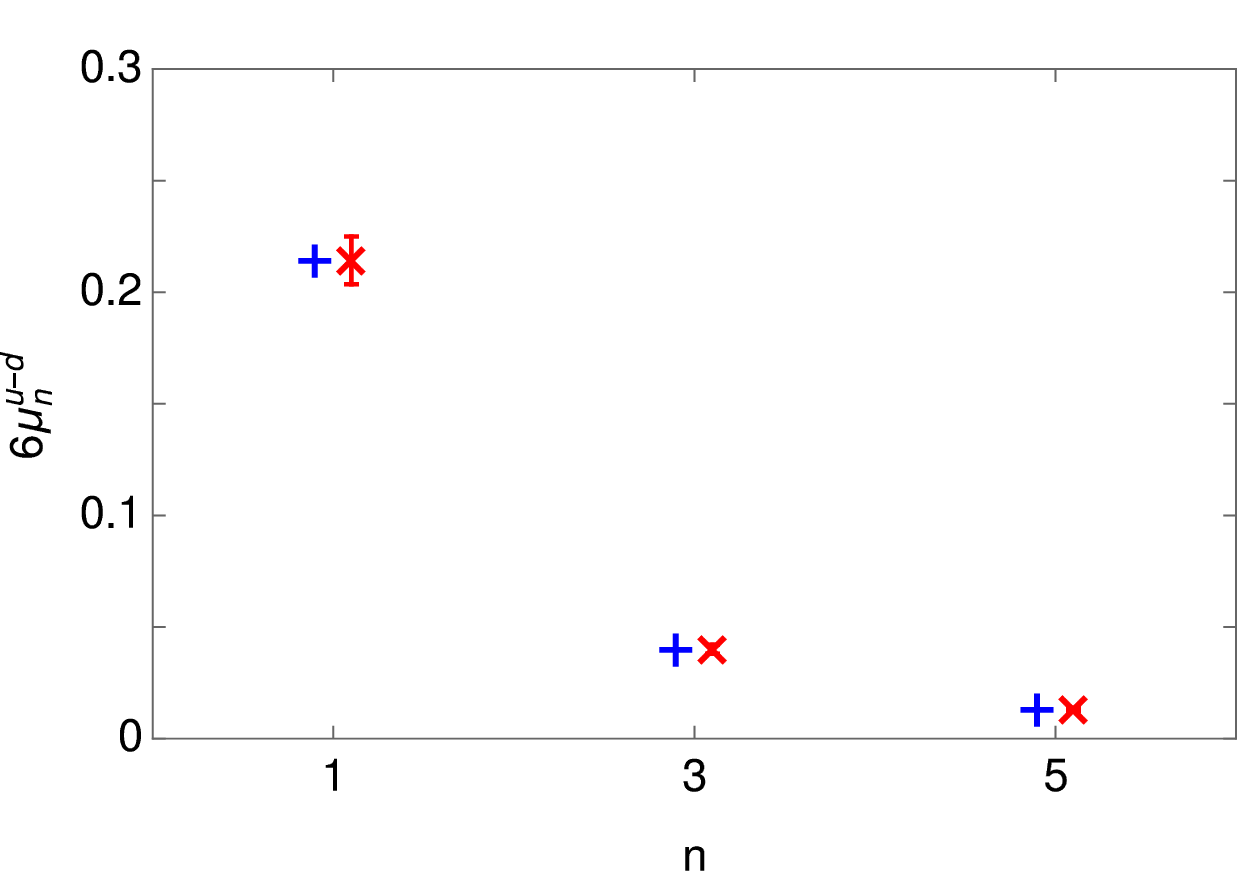,width=10cm,clip=}
\end{center}
\vspace*{-0.75cm}
\caption{The lowest three moments $\mu_n^{u-d}$ with $n=1, 3$ and $5$. The blue pluses ($\textcolor{blue}{+}$) are the target numbers, $6\,\mu_n=\int_0^1 dx x^n \left[u(x)-d(x)\right]$. The red crosses ($\textcolor{red}{\times}$) are the solution of the OPE (\ref{va}). For the error estimate see the text. At $n=3,5$ the error is smaller than the symbol.}
\label{fig3}
\end{figure}

The mathematical methods described above enable us to obtain the lower moments of the structure function, as well as $F_1(x)$ as a whole, rather accurately from a relatively small set of values of $T_{33}(\omega)$. To demonstrate that, we start from the experimental nonsinglet structure function $F_1^{u-d}(x)$, which we parameterize by 
\begin{equation}
6\hspace*{0.2mm} x\hspace*{0.25mm} F_1^{u-d}(x) = x\left[u(x)-d(x)\right] 
\label{mstw}
\end{equation}
with $u(x)$ and $d(x)$ taken to be the LO parton distributions at the scale $q^2=1\,\mbox{GeV}^2$~\cite{Martin:2009iq}. From (\ref{mstw}) we compute $T_{33}(\omega)$ for $\omega=0.1$, $0.2$, $\cdots$, $0.9$ using (\ref{opess2}). 
The result is shown in Fig.~\ref{fig2}. This we consider our `data', from which we want to retrieve the moments $\mu$ and structure function $F_1^{u-d}(x)$, and to which the lattice results (see Fig.~\ref{fig6}) will aspire to. 

\begin{figure}[!t]
\begin{center}
\epsfig{file=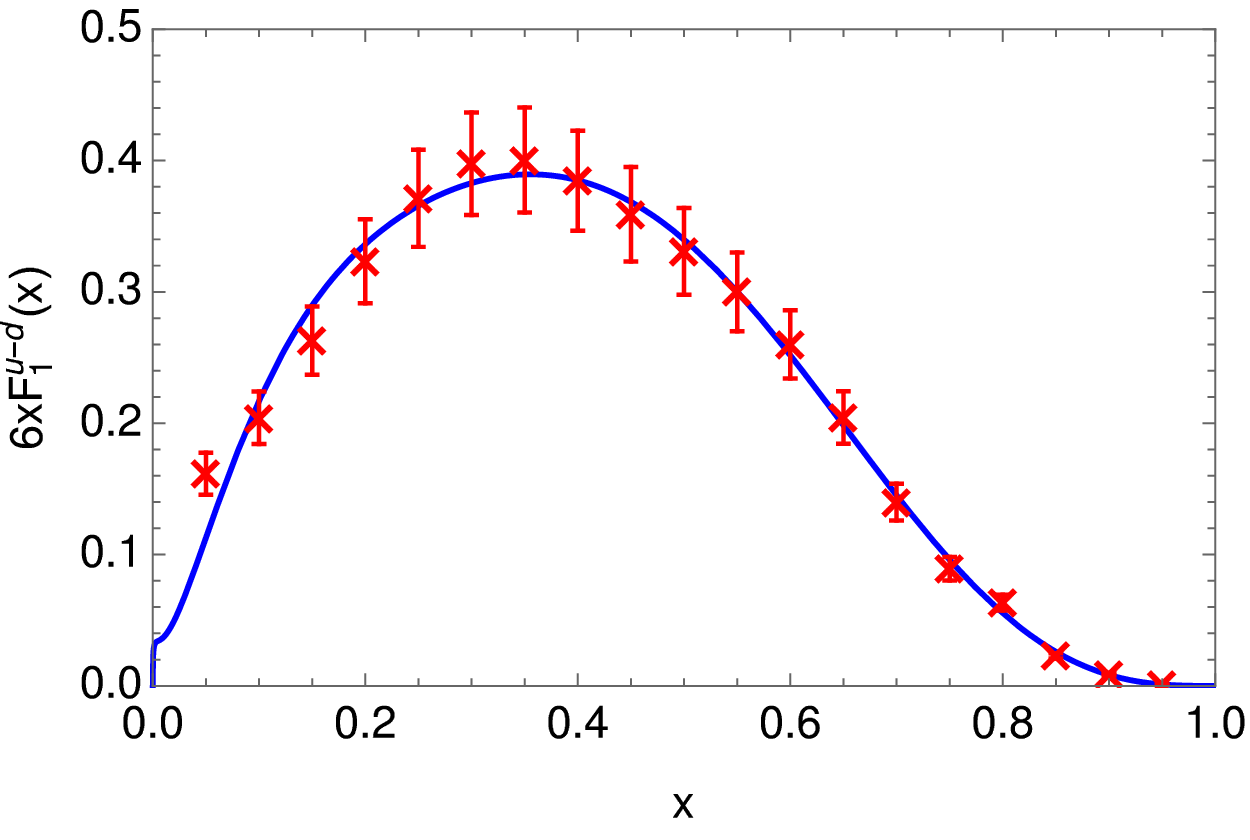,width=10cm,clip=}
\end{center}
\vspace*{-0.75cm}
\caption{The structure function $F_1^{u-d}(x)$. The solid line is the target structure function, $\displaystyle 6xF_1(x) = x \left[u(x)-d(x)\right]$. The crosses ($\textcolor{red}{\times}$) are the solution of the SVD (\ref{svd}). For the error estimate see the text.}
\label{fig4}
\end{figure}

To compute the moments, we fit the interpolating polynomial (\ref{poly}) to the lowest five `data' points $t_1, \cdots, t_5$ with $M = 5$, which is equivalent to solving the set of linear equations (\ref{va}) for $\mu_1, \mu_3, \cdots, \mu_9$. The result is shown in Fig.~\ref{fig3} for the lowest three moments, together with the target numbers. The first three moments are accurately reproduced.  

\begin{figure}[!b]
\begin{center}
\epsfig{file=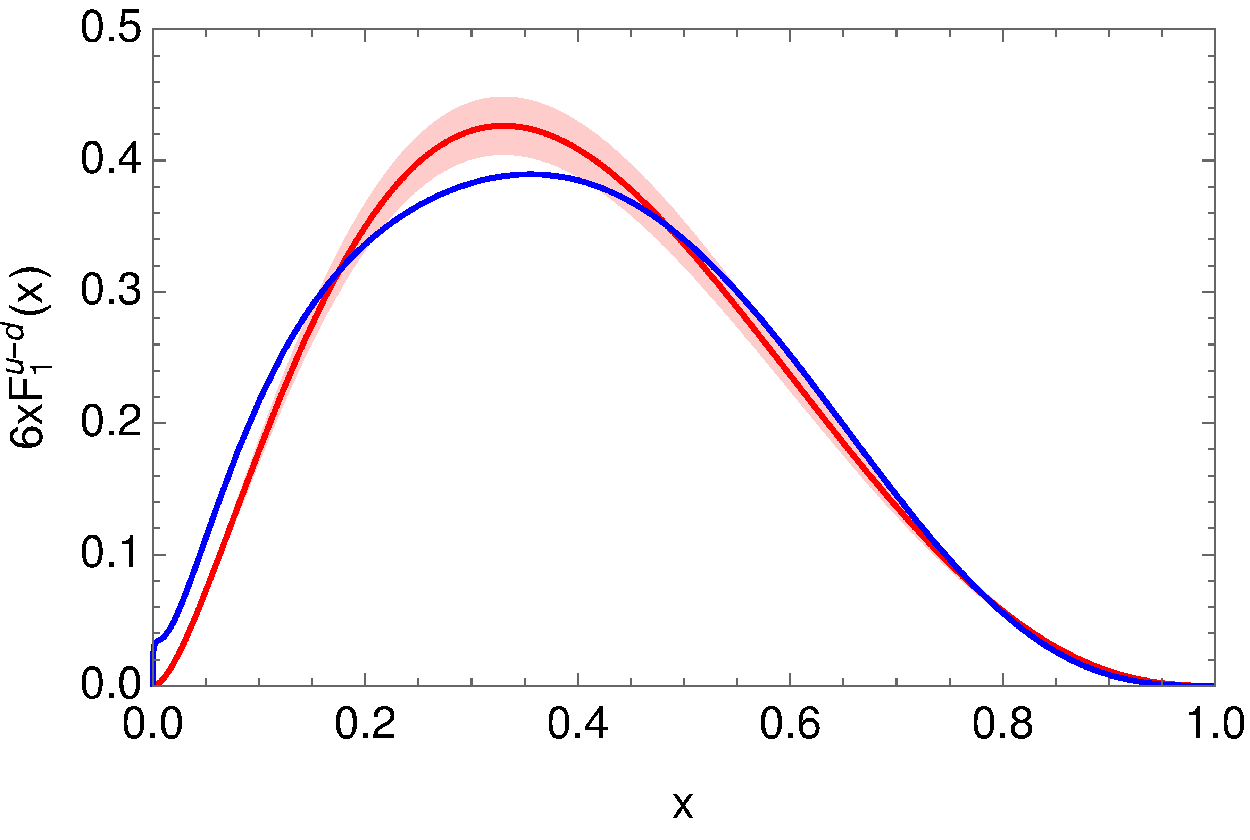,width=10cm,clip=}
\end{center}
\vspace*{-0.75cm}
\caption{The structure function $F_1^{u-d}(x)$ obtained from the Mellin transform of (\ref{me}) fitted to the moments (\textcolor{red}{$-$}), compared with the target structure function $\displaystyle 6xF_1(x) = x \left[u(x)-d(x)\right]$ (\textcolor{blue}{$-$}). For the error estimate see the text.}
\label{fig5}
\end{figure}

To retrieve the structure function $F_1^{u-d}(x)$, we apply the SVD (\ref{svd}) to our `data'~\cite{ma}. Three out of nine eigenvalues of $W$ turned out to be zero. In Fig.~\ref{fig4} we show the result for $M = 19$. It shows that the structure function $F_1^{u-d}$ can be well reproduced from a relatively small set of data, except perhaps for $x \lesssim 0.05$. Similar results are obtained for the singlet structure function $F_1^{u+d+\bar{u}+\bar{d}+\bar{s}}$.
We have not made any attempts to optimize the SVD. It can be improved in several respects. A Bayesian approach~\cite{Liu:2016djw} to alleviate overfitting, for example, might lead to particularly robust results.  

There are other possibilities as well to compute the structure function from the Compton amplitude. A particularly promising approach is to fit the moments, for example in the interpolating polynomial (\ref{poly}), by an appropriate function $\mu(s)$ with $\mu(n)=\mu_n$ and employ an inverse Mellin transform on $\mu(s)$ to obtain $F_1(x)$. It turns out that the moments can be fitted surprisingly well by the simple expression
\begin{equation}
\mu(s)=A\,(s+\alpha)^{-\beta}\,,
\label{me}
\end{equation}
for which the inverse Mellin transform is known analytically~\cite{er}. Starting from the moments $6\,\mu_n=\int_0^1 dx x^n \left[u(x)-d(x)\right]$, the result of the Mellin transform is shown in Fig.~\ref{fig5}. 


The analysis so far has been limited to $\omega \in[0,1]$. The SVD method can be extended to larger values $\omega > 1$ without problem. This will allow us to probe the small-$x$ region of $F_1(x)$, which is not accessible through moments of the structure function. Indeed, by extending the calculation to $\omega = 2$, we were able to retrieve the singlet structure function $F_1^{u+d+\bar{u}+\bar{d}+\bar{s}}(x)$~\cite{Martin:2009iq} down to fractional momenta $x \lesssim 0.005$, which was not possible before. Odd moments of the structure functions can be obtained by also including the local axial vector current $\bar{\psi}_f(x)\gamma_3 \gamma_5 \psi_f(x)$ to (\ref{add}) and studying the interference with the vector current. This is achievable through a simple extension of the procedure described above. The method can be generalized to nonforward Compton scattering as well. That will allow us to derive generalized parton distribution functions (GPDs).  

\begin{figure}[!b]
\begin{center}
\epsfig{file=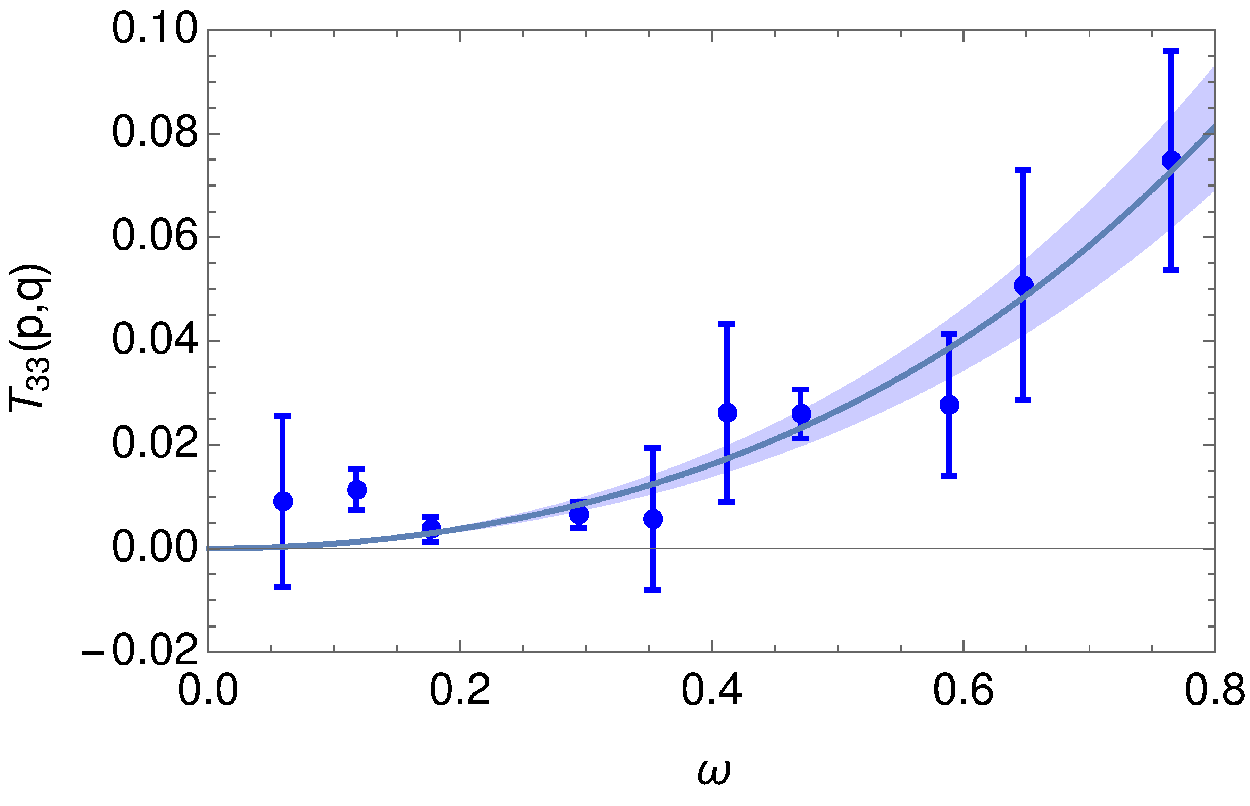,width=10cm,clip=}
\end{center}
\vspace*{-0.75cm}
\caption{The proton Compton amplitude $T_{33}(p,q)$ for momenta $\vec{p}=(2,-1,0)$, $(-1,1,0)$, $(1,0,0)$, $(0,1,0)$, $(2,0,0)$, $(-1,2,0)$, $(1,1,0)$, $(0,2,0)$, $(2,1,0)$, $(1,2,0)$, from left to right, and $\vec{q}=(3,5,0)$, in lattice units. The current has been attached to the $d$ quark, leading to the `handbag' diagram in Fig.~\ref{fig1}. $Z_V$ has been taken from~\cite{Constantinou:2014fka}. The solid line shows a sixth order polynomial fit (giving $\chi^2/{\rm dof}=0.9$), and the shaded area shows the error.}
\label{fig6}
\end{figure}

There is the question what accuracy can be achieved with real data. It turns out that the second derivative of the nucleon energy can be computed rather accurately. In a proof-of-principle study we have computed (\ref{fh}) from $O(900)$ configurations generated at the SU(3) symmetric point~\cite{Bietenholz:2011qq} on a $32^3\times 64$ lattice with lattice spacing $a\approx 0.074\,\mbox{fm}$. First results are presented in Fig.~\ref{fig6}, where the contribution from  $\displaystyle \mathcal{F}_1(0,q^2) = \left.- 2\,E_\lambda(0,q)\, \partial^2 E_\lambda(0,q)/\partial \lambda^2\,\right|_{\lambda=0}$ has been subtracted. The precision for lattice momenta $\vec{p}^2=1$ and $2$ is already quite impressive. We should be able to improve on the precision of the data at higher momenta by employing `momentum smearing' techniques~\cite{Bali:2016lva}, which has not been attempted here. Based on this result, we consider an overall projected error of $10\%$ on the Compton amplitude $T_{33}$, marked by the shaded area in Fig.~\ref{fig2}, a conservative estimate. The resulting errors on $\mu_n$ and $F_1^{u-d}(x)$, shown in Figs.~\ref{fig3}, \ref{fig4} and \ref{fig5}, have been obtained from replacing the initial values of $T_{33}(\omega_i)$ by the corresponding numbers on the confidence envelopes. This leads us to conclude that the entire structure function, including its moments, can be reconstructed from a lattice calculation of the Compton amplitude with unprecedented accuracy, devoid of any renormalization and mixing issues. 



\section*{Acknowledgement}

GS thanks Akaki Rusetsky for useful discussions. The numerical calculations were carried out on the IBM BlueGene/Qs at Edinburgh and J\"ulich using DIRAC 2 and NIC resources, on the Cray XC30 at HLRN, Berlin and Hannover, and on the NCI National Facility at Canberra. HP and GS are supported by DFG Grant Nos.\ SCHI 422/10-1 and SCHI 179/8-1. PELR is supported by the STFC under contract ST/G00062X/1. RDY and JMZ are supported by the Australian Research Council Grant Nos.\ FT120100821, FT100100005 and DP140103067.




\begin{thebibliography}{99}

\bibitem{Martinelli:1996pk}
  G.~Martinelli and C.T.~Sachrajda,
  Nucl.\ Phys.\ B {\bf 478} (1996) 660
  [hep-ph/9605336].

\bibitem{dislat}
For pioneering work see:
  G.~Martinelli and C.T.~Sachrajda,
  Nucl.\ Phys.\ B {\bf 316} (1989) 355,
  M.~G\"ockeler, R.~Horsley, E.-M.~Ilgenfritz, H.~Perlt, P.E.L.~Rakow, G.~Schierholz and A.~Schiller,
  Phys.\ Rev.\ D {\bf 53} (1996) 2317
  [hep-lat/9508004].\\ 
For recent reviews of the subject see:  M.~Constantinou,
  PoS CD {\bf 15} (2015) 009
  [arXiv:1511.00214 [hep-lat]];
  H.-W.~Lin,  
  arXiv:1612.09366 [hep-lat]; and references cited therein.

\bibitem{Martinelli:1998hz}
  G.~Martinelli,
  Nucl.\ Phys.\ Proc.\ Suppl.\  {\bf 73} (1999) 58
  [hep-lat/9810013].

\bibitem{op}
  S.~Capitani, M.~G\"ockeler, R.~Horsley, H.~Oelrich, D.~Petters, P.E.L.~Rakow and G.~Schierholz,
  Nucl.\ Phys.\ Proc.\ Suppl.\  {\bf 73} (1999) 288
  [hep-lat/9809171];
  S.~Capitani, M.~G\"ockeler, R.~Horsley, D.~Petters, D.~Pleiter, P.E.L.~Rakow and G.~Schierholz,
  Nucl.\ Phys.\ Proc.\ Suppl.\  {\bf 79} (1999) 173
  [hep-ph/9906320];
  W.~Bietenholz, N.~Cundy, M.~G\"ockeler, R.~Horsley, H.~Perlt, D.~Pleiter, P.E.L.~Rakow, G.~Schierholz, A.~Schiller, T.~Streuer and J.M.~Zanotti,
  PoS LAT {\bf 2009} (2009) 138
  [arXiv:0910.2437 [hep-lat]].

\bibitem{Caracciolo} S.~Caracciolo, A.~Montanari and A.~Pelissetto,
Nucl. Phys. B (Proc.~Suppl.) {\bf 73} (1999) 273, ibid.\ {\bf 83-84} (2000) 875, JHEP {\bf 9} (2000) 45;
 G.C.~Rossi, Chin.\ J.\ Phys.\ {\bf 38} (2000) 721.

\bibitem{Detmold:2005gg}
  W.~Detmold and C.J.D.~Lin,
  Phys.\ Rev.\ D {\bf 73} (2006) 014501
  [hep-lat/0507007].

\bibitem{dis}
  R.~Devenish and A.~Cooper-Sarkar, {\it Deep Inelastic Scattering}, Oxford   University Press (2003, Oxford, UK);
  A.V.~Manohar,
  in {\it Lake Louise 1992, Symmetry and Spin in the Standard Model}, p.\ 1-46
  [hep-ph/9204208];  K.-F.~Liu,
  Phys.\ Rev.\ D {\bf 62} (2000) 074501 [hep-ph/9910306].

\bibitem{Agadjanov:2016cjc}
  See, for example: A.~Agadjanov, U.-G.~Mei{\ss}ner and A.~Rusetsky,
  arXiv:1610.05545 [hep-lat].

\bibitem{Horsley:2012pz}
 R.~Horsley, R.~Millo, Y.~Nakamura, H.~Perlt, D.~Pleiter, P.E.L.~Rakow, G.~Schierholz, A.~Schiller, F.~Winter, J.M.~Zanotti,
  Phys.\ Lett.\ B {\bf 714} (2012) 312
  [arXiv:1205.6410 [hep-lat]];
  A.J.~Chambers, R.~Horsley, Y.~Nakamura, H.~Perlt, D.~Pleiter, P.E.L.~Rakow, G.~Schierholz, A.~Schiller, H.~St\"uben, R.D.~Young and J.M.~Zanotti,
  Phys.\ Rev.\ D {\bf 90} (2014) 014510
  [arXiv:1405.3019 [hep-lat]];
A.~J.~Chambers, R.~Horsley, Y.~Nakamura, H.~Perlt, D.~Pleiter, P.E.L.~Rakow, G.~Schierholz, A.~Schiller, H.~St\"uben, R.D.~Young and J.M.~Zanotti,
  Phys.\ Rev.\ D {\bf 92} (2015) no.11,  114517
  [arXiv:1508.06856 [hep-lat]]; 
A.J.~Chambers, J.~Dragos, R.~Horsley, Y.~Nakamura, H.~Perlt, D.~Pleiter, P.E.L.~Rakow, G.~Schierholz, A.~Schiller, K.~Somfleth, H.~St\"uben, R.D.~Young and J.M.~Zanotti,
  arXiv:1702.01513 [hep-lat].

\bibitem{Bakeyev:2003ff}
  T.~Bakeyev, M.~G\"ockeler, R.~Horsley, D.~Pleiter, P.E.L.~Rakow, G.~Schierholz and H. St\"uben,
  Phys.\ Lett.\ B {\bf 580} (2004) 197
  [hep-lat/0305014].

\bibitem{nr}
W.H. Press, B.P. Flannery, S.A. Teukolsky and W.T. Vetterling, {\it Numerical Recipes}, Cambridge University Press (1989, Cambridge, UK).

\bibitem{ma} Wolfram Research, Inc., {\it Mathematica}, Version 11.0 (2016, Champaign, USA).

\bibitem{Martin:2009iq}
  A.D.~Martin, W.J.~Stirling, R.S.~Thorne and G.~Watt,
  Eur.\ Phys.\ J.\ C {\bf 63} (2009) 189
  [arXiv:0901.0002 [hep-ph]].

\bibitem{Liu:2016djw}
  K.-F.~Liu,
  PoS LATTICE {\bf 2015} (2016) 115
  [arXiv:1603.07352 [hep-ph]].

\bibitem{er}
A. Erd\'{e}lyi, {\it Table of Integral Transforms}, Vol. I and II, McGraw-Hill (1954, New York, USA).

\bibitem{Bietenholz:2011qq}
  W.~Bietenholz, V.~Bornyakov, M.~G\"ockeler, R.~Horsley, W.G.~Lockhart, Y.~Nakamura, H.~Perlt, D.~Pleiter, P.E.L.~Rakow, G.~Schierholz, A.~Schiller, T.~Streuer, H.~St\"uben, F.~Winter and J.M.~Zanotti,
  Phys.\ Rev.\ D {\bf 84} (2011) 054509
  [arXiv:1102.5300 [hep-lat]].

\bibitem{Constantinou:2014fka}
  M.~Constantinou, R.~Horsley, H.~Panagopoulos, H.~Perlt, P.E.L.~Rakow, G.~Schierholz, A.~Schiller and J.M.~Zanotti,
  Phys.\ Rev.\ D {\bf 91} (2015) no.1,  014502
  [arXiv:1408.6047 [hep-lat]].

\bibitem{Bali:2016lva}
  G.~S.~Bali, B.~Lang, B.~U.~Musch and A.~Sch\"afer,
  Phys.\ Rev.\ D {\bf 93} (2016) no.9,  094515
  [arXiv:1602.05525 [hep-lat]].




\end{thebibliography}
\end{document}